\documentclass[12pt]{article}

\begin{document}

\begin{center}
{\large {\bf
ALPHA-PARTICLE CONDENSATE IN NUCLEAR MATTER AT NORMAL DENSITY AND
STATISTICS OF COMPOSITE BOSONS}}
\end{center}

\begin{center}
{\large I.A.Gnilozub$^1$, S.D.Kurgalin$^2$, Yu.M.Tchuvil'sky $^{1,3}$} \\

{\it $^1$ Moscow State University,
119992 Moscow, Russia}
{$^2$ \it Voronezh State University, 394693 Voronezh, Russia}
{\it $^3$ Justus-Liebig-Universitat, D-35392 Giessen, Germany}
\end{center}

{\small It is proved that $\alpha$-particle states are well described by the
Elliott SU(3) model. This result is used to analyze the alpha-particle
condensation effect. It is shown that these states possess the basic attributes
of the $\alpha$-condensate and, also, the normal nuclear density on frequent
occasions. The statistics of $\alpha$-particles (and of arbitrary composite
bosons) turns out to be something other than the Bose-Einstein, Fermi-Dirac
statistics, and parastatistics.}

\vspace{5mm}

At present there is keen interest in search for a manifestation of the
$\alpha$-particle condensate in nuclear matter \cite{Rop,TH,SH}. Both the
possibility of its existence and general properties are discussed. The specific
problem of nuclear physics, namely, search for a new line of states as well as
the general physical problem of formation and condensation of composite bosons
in an arbitrary fermion system are promising areas of application of the alpha
particle condensate (APC) concept. The main theoretical approach to the
problems is analysis of $\alpha$-condensed matter at low densities of a nuclear
system \cite{Rop}. The closely resembling possibility is to search for a
condensed state near the $k\alpha$-particle $(k=A/4$) disintegration threshold
\cite{TH,SH} of an A/2=N=Z-even nucleus. Solution of the Hill-Wheeler equation
for the size parameters of a $k\alpha$-particle system is proposed as a
definition of the $\alpha$-condensate. It is shown that there are such
solutions in $^{12}$C, $^{16}$O and heavier nuclei. Their density is also low.
Experimental spectra of these nuclei contain levels which can be considered as
candidates for the states of the type discussed. Evidently these states display
the properties of multi alpha particle states (MAPS) in an arbitrary nuclear
collision.

At the same time there is another possibility to construct MAPS which exhibit
analogous properties in the same processes \cite{KT}. This procedure will be
followed in the present paper. We analyze the possibility to reduce the
A-nucleon problem to the $k\alpha$-particle one and demonstrate that for
definite states of an A/2=N=Z-even nucleus the Hamiltonian of the well-known
Elliott model \cite{El} allows one to do it precisely. In other words, in some
states such a nucleus is completely described by the $\alpha$-particle
dynamics, and a system in such a state behaves very much like that discussed in
\cite{TH}. In addition its wave functions (WFs) can be written in a similar
form. So the idea to consider these states as $\alpha$-condensate appears to be
reasonable.

The ground and low-lying states of light N=Z-even nuclei are well described by
the Elliott model. Therefore these states are also analyzed.

The key point of the present work is the question of wether $\alpha$-particles
can occupy one and the same level and what this level is. To shed light on this
problem selection rules of MAPS are investigated. On this basis the statistics
of $\alpha$-particles in a nuclear system is revealed. The difference between
the manifestation of a quasiboson condensate in nuclei (dense systems of the
relatively small fermion number) and in larger-volume many-fermion objects is
discussed.

The approach proposed for the declared purposes is based on the microscopic
(operating in the space of A-nucleon wave functions (WFs)) SU(3) Hamiltonian
\cite{El} (generalized by incorporating the $Q^3$, $QL^2$ terms) which is
constructed from a number of the commuting invariant operators of the SU(3)
group and the related subgroups:

\begin{equation}
H = H_{osc.} + f_1(\hat L^2) + f_2(QQ) + f_3((Q\otimes Q)Q) +
f_4(Q(\hat L\otimes \hat L)), \label{Ham}
\end{equation}
where $H_{osc.}$ is the oscillator Hamiltonian; $\hat L$, the operator of
angular momentum, and $Q$  is the quadripol operator. Its component has the
form:

\begin{equation}
Q^m = \sqrt{4\pi /5}\sum_i((\rho_i^2/\rho_{i_0}^2) Y_{2m}(\theta_{\rho_i},
\phi_{\rho_i}) + (p_i^2/p_{i_0}^2) Y_{2m}(\theta_{p_i},\phi_{p_i}))
\end{equation}
and $\rho_i$ is the Jacobi coordinate, and $p_i$ is the operator of related
linear momentum.

The eigenvalues of the Hamiltonian are expressed as:

$$E = N\hbar \omega + b_1L(L+1) + b_2(2/3)(\lambda^2 + \mu^2 +
\lambda \mu +3(\lambda+\mu)) + $$
$$b_3(1/9)(\lambda-\mu)[(\lambda+2\mu)
(2\lambda+\mu) + 9(\lambda+\mu+1) - $$
\begin{equation}
(\lambda^2 + \mu^2 +
\lambda \mu +3(\lambda+\mu))] +
b_4\Omega, \label{EV}
\end{equation}
where $\omega$ is the oscillator frequency and $\Omega$ is the eigenvalue of
the Bargman operator $\hat \Omega$ \cite{BM}. Eigenfunctions of the Hamiltonian
$\Psi_A$ must be characterized by the following global (related to the system
as a whole) quantum numbers (QNs): the principal QN $N= \sum_j n_j$
($n_j=2\tilde n_j + l_j$ is the principal QN of the one-nucleon wave function
(WF); $\tilde n_j$, number of nodes in it, and $l_j$ is its angular momentum),
the Elliott symbol $(\lambda \mu)$ characterizing the SU(3) irreducible
representation, angular momentum $L$ and the value $\Omega$. Naturally the QNs
which describe spin-isospin wave functions are also necessary. The
eigenfunctions can be written in various forms, e.g. using the oscillator shell
model as the most universal approach. The translationally invariant version of
the model \cite{KSSE}, free of the redundant center-of-mass (CM) coordinate of
the system and of spurious states is preferable. Sometimes the  eigenstates
turn out to be degenerate with certain multiplicity and so some additional QNs
$\eta$ which characterize $\Psi_A$ may be required for a complete
classification. The ground and other low-lying states of nuclei with the
maximum allowed value of $\lambda$ and $N \leq N_{min}+1$ in the Elliott model
are not degenerate.

The principal point of the approach is that in some cases these eigenfunctions
can be expressed equivalently in the shell model and in the multicluster form
with the same quantum numbers. For the N=Z even nucleus possible clusters are
$\alpha$-particles:

\begin{equation}
\Psi_\Delta ^A \equiv \hat A \hat N^{-1/2} \prod_{i=1}^k \Psi_{\alpha_i}\cdot
\Psi_{\tilde \Delta}(\{\rho_k\}) S T. \label{Fun}
\end{equation}
Here  $\Psi_{\tilde \Delta}(\{\rho_k\})$ is  the WF of the  relative motion of
$k$ $\alpha $-clusters in the SU(3) scheme; $\{\rho_k\}$, the set of Jacobi
coordinates, $\Delta $ contains both the unambiguously determined Young frame
$[f]=[4^{A/4}]$, spin $S=0$, isospin $T=0$ and the spatial QNs discussed above,
$\tilde \Delta $ denotes the same number as $ \Delta $ except for $S$ and $T$,
and $\hat A$ is the antisymmetrizer. The operator

\begin{equation}
\hat N \equiv <\hat A  \prod_{i=1}^k \Psi_{\alpha_i} \delta (\{\rho_k - \rho'_k\})
| \hat A  \prod_{i=1}^k \Psi_{\alpha_i} \delta (\{\rho_k - \rho''_k\})  > - \label{Norm}
\end{equation}
is the multicluster overlap kernel of the resonating group model. Its
eigenfunctions are also characterized by the QNs $N$, $(\lambda \mu)$, $L$ and
$\Omega $. Consequently the WF $\Psi_{\tilde \Delta } (\{\rho_k\})$ can be
constructed as the eigenfunction of the kernel $\hat N$ related to the
eigenvalue $\epsilon_{\tilde \Delta}$. In other words this function is
"selfreproductive", i.e. it does not change its form under the transformation:

\begin{equation}
\hat O \hat A \prod_{i=1}^k \Psi_{\alpha_i}\cdot
\Psi_{\tilde \Delta}(\{\rho_k\}) S T = \epsilon_{\tilde \Delta }
\Psi_{\tilde \Delta } (\{\rho_k\}). \label{Self}
\end{equation}
Here $\hat O \equiv \int d\{\xi_{\alpha_i}\} d\{\rho'_k\} \prod_{i=1}^k
\Psi_{\alpha_i} \delta (\{\rho_k - \rho'_k\}) $; the $\{\xi_{\alpha_i}\}$ is
the set of the internal coordinates of all $\alpha$-particles, and
$\epsilon_{\tilde \Delta }$ is the eigenvalue of the kernel $\hat N$. The
function $\Psi_{\tilde \Delta } (\{\rho_k\})$ is  symmetric about a permutation
of $\alpha$-particles. In case of the multiplicity of eigenstates of the
Hamiltonian greater than unity the eigenvalues of the kernel can be used for
defining additional QNs $\eta$  in $\tilde \Delta$ and $\Delta$.

In fact this formalism is a MAP generalization of the two $\alpha$-particle
representation of the $^8$Be WF \cite{WK}. Naturally the relation (\ref{Fun})
cannot be written using the WFs of different constituents except for
$\alpha$-particles (a short list of states which are exclusions involves
$^{16}$O or $^{40}$Ca as constituents) and it is valid only for the indicated
type of Young frames.

If one excludes forbidden states annihilated by the antisymmetrizer then both
the full-space SU(3) Hamiltonian (\ref{Ham}) and reduced Hamiltonian of the
same form operating in the space of the relative WFs $\Psi_{\tilde
\Delta}(\{\rho_k\}) $ are equivalent, i.e., they lead to the same spectra
(\ref{EV}). So in such a dynamics $\alpha$-particles can be consider as
structureless constituents and the N=Z-even nuclear system described by the
Hamiltonian (\ref{Ham}) behaves as a system of $k$ stable $\alpha$-particles.
This property of the states (\ref{Fun}) is necessary for a model of APC. The
most probable response of such a system to an external impact is its
disintegration into $\alpha$-particles and/or larger MAP parts. In this respect
there is in fact no difference between the discussed state and the
$\alpha$-condensed one defined in \cite{TH}.

It should be noted that the Elliott model provides a rather good description
for the ground and low-lying states of the N=Z-even nuclei of the 1p-,
(2s-1d)-shell, and of the start of the  (2p-1f)-shell \cite{Harv}. So the
states of such nuclei with $[f]=[4^{A/4}]$ are good examples of MAPS. An
extended version of the model \cite{GKT} also explains various qualitative
properties and provides rather good quantitative description of highly excited
$\alpha$-particle states populated in elastic $\alpha$-scattering \cite{Gold}
and in $\alpha$-transfer reactions. These states are a particular case of MAPS
characterized by the SU(3) representations $(\lambda \mu) \subset (\lambda_t
\mu_t)\times (n0)$, where the second and the last Elliott symbols are related
to the ground state of a target nucleus and to the relative motion of the
target and $\alpha$-particle, respectively. Analogous restrictions are imposed
on the angular momentum of an $\alpha$-particle state. Such states are
numerous. Consequently, there are many MAPS, too. Thus the properties of APC
may in some cases be an attribute of nuclear matter at the normal density.

For a further analysis of the properties of MAPS the most important question is
whether $\alpha$-particles of such a system can occupy one and the same
$\alpha$-particle level and what this level is. Elucidation of this problem
requires an analysis of the occupation of one-$\alpha$-particle levels. For
this purpose it is convenient to pass from the translationally invariant wave
function (\ref{Fun}) to its ordinary shell model analogue:

\begin{equation}
\Psi _A^{(SM)}=\Psi _A \Phi_{000}(R_{cm}) \label{SM},
\end{equation}
where the last function is the WF of zero oscillations of the CM of the system.
This function is commutative with the antisymmetrizer and the overlap kernel.
Thus by analogy with the function (\ref{Fun}) the properties of the function
(\ref{SM}) are determined by the properties of the product $\Psi_{\tilde
\Delta}(\{\rho_k\}) \Phi_{000}(R_{cm})$. This function is selfreproducing and
symmetric because both multipliers are also symmetric. Therefore one can
rewrite this product as a symmetric superposition of the following products of
one-alpha-particle WFs with a proper scheme of coupling of their partial QNs
into the global quantum number $\tilde \Delta \equiv [f] N (\lambda \mu) \Omega
L \eta '$:

\begin{equation}
\Psi_{\tilde \Delta}(\{R_i\}) \equiv \prod_{i=1}^k \varphi_{(N_i0)}(R_i):\{(\lambda_i \mu_i) \}
[4^{A/4}] N \Omega L, \label{RSM}
\end{equation}
where $N_i$ is the principle QN of the one $\alpha$-particle WF
$\varphi_{(N_i0)}(R_i)$, $N = \sum_{i=1}^{A/4} N_i $; the $(N_i0)$, the SU(3)
representation of three-dimention one-particle motion, and $R_i$ is the
coordinate of the center of mass of an $\alpha$-particle. The intermediate
Elliott symbols $(\lambda_1 \mu_1)=(N_i 0)$, ..., $(\lambda_i \mu_i) \subset
(\lambda_{i-1} \mu_{i-1}) \times (N_i 0)$, ..., $(\lambda_k \mu_k) \equiv
(\lambda \mu)$ determine the SU(3) coupling scheme (there is no multiple
representation in such products of the SU(3) representations) and serve as
additional QNs $\eta '$. Using the SU(3) coupling scheme is necessary because
partial (one-particle) angular momenta cannot be determined simultaneously with
$(\lambda \mu)$. Obviously this global Elliott symbol and Young frame must be
compatible. Operation of the antisymmetrizer involved in $\Psi _A^{(SM)} $
determines selection rules for QNs ${\tilde \Delta}$ of the WFs which are not
annihilated by $\hat A$. The selection rule for the principal QN $N$ is a
trivial result of the oscillator shell model. Namely, for $^4$He, $^8$Be
$^{12}$C $^{16}$O, $^{20}$Ne, $^{24}$Mg, ..., $^{40}$Ca, $^{44}$Ti nuclei the
lower limit of $N$ is $N_{min}=0, 4, 8, 12, 20, 28, ..., 60, 72$ etc. The basis
of the functions (\ref{RSM}) is complete in the space of the products of
one-$\alpha$-particle WFs, therefore this limiting condition is valid for
arbitrary components of multi alpha particle WFs but not only for the
eigenfunctions of the Hamiltonian (\ref{Ham}). Each component (\ref{RSM}) of an
arbitrary MAPS is characterized by certain occupation numbers of
$\alpha$-particles at the levels $E_i = (N_i +3/2) \hbar \omega$. For $A>4$ the
conventional condensed state $(0s)^{A/4}$ is forbidden by the Pauli principle,
therefore the rigorous definition of APC cannot be satisfied in principle. Even
the presented initial stage of consideration demonstrates unusual properties of
MAPS.

So we concentrate on the statistics of $\alpha$-particles. In order to make it
more pronounced it is convenient to construct the WFs $\Psi _A^{(SM)}$ in a
different way. Indeed one can write:

\begin{equation}
\Psi _A^{(SM)}=\hat A \prod_{i=1}^k \Psi_{\alpha_i}\cdot
\hat P \Psi_{\tilde \Delta} (\{R_i\}) \label{C1}
\end{equation}
using an arbitrary component from (\ref{RSM}) which is not annihilated by the
operators $\hat A$ and $\hat P$ . Here $\hat P$ is the projection operator on
the states of the type $\Psi^{rel}(\{\rho_k\}) \Phi_{000}(R_{cm})$, i.e. the
states which are related to zero oscillation of the CM. The explicit expression
of this operator can be found in \cite{FGZ}. The spatial WF on the right hand
side of (\ref{C1}) is no longer selfreproducing and, in general, is not
symmetric. Nevertheless the WF on the left hand side is symmetric and, what is
more, for non-degenerate states it coincides, when normalized, with the
function (\ref{SM}). For degenerate states it turns out to be a superposition
of the functions (\ref{SM}) (summation over $\eta$).

In order to prove that any component survives under the operation $\hat A$ one
can expand the function $\prod_{i=1}^k \Psi_{\alpha_i}\cdot \hat P \Psi_{\tilde
\Delta}(\{R_i\})$ onto a superposition of the products of the shell-model
one-nucleon wave functions. Even the appearance of one term with an allowed
configuration of nucleons is a sufficient condition of the survival.

According to the selection rules presented a construction procedure of the
component (\ref{RSM}) for the $k\alpha$-particle system is as follows. The
first $\alpha$-particle occupies the $N_i = 0$ level (the $(N_i0)=(00)^1$
$\alpha$-particle configuration in the SU(3) notation, being the ground state
of the $^4$He nucleus in the oscillator shell model). The second one can occupy
any level with an even value of $N_i \geq 4$. If it is placed on the $N_i=4$
level (the $(00)^1(40)^1$ configuration being the ground state band of $^8$Be)
then the third $\alpha$-particle can occupy the same level (the $(00)^1(40)^2$
configuration being the ground state band of $^{12}$C) and higher levels. The
$\alpha$-particle over the ground state of $^{12}$C can occupy the levels with
$N_i \geq 4$ with the exclusion of $N_i=5$ because this state is spurious i.e.
it is annihilated by the operator $\hat P$. For the fifth $\alpha$-particle
over the ground state of $^{16}$O ($(00)^1(40)^3$) all levels with $N_i \geq 8$
are allowed. The global $(\lambda \mu)$ and intermediate $(\lambda_i \mu_i)$
Elliott symbols are constructed as demonstrated above. They determine the
collective properties of MAPS. For brevity we omit the discussion about their
selection rules. The selection rules for the quantities $ \Omega$ and $L$,
which determine rotational bands, follow immediately from the
SU(3)$\supset$O(3) reduction chain.

For the fixed number of $\alpha$-particles $k=A/4$ and for the principal QN $N$
which satisfies the conditions under discussion it is also possible to choose
the number of values $\{N_i\}$ in the form: $N_i= [N/k]$ (the $[N/k]$ is the
integer part of the fraction $N/k$) for $i \leq k[N/k] - N +k$, and $N_i =[N/k]
+ 1$ for the higher values of $i$. For an integer value of $N/k$ all
$\alpha$-particles can occupy one and the same level with $N_i= N/k$ (certainly
if lower ones are free). As an important example it is allowed for the ground
states of an oscillator magic nuclei to locate all
$k_{mag}=A_{mag}/4=(N_{i^{mag}}+3)(N_{i^{mag}}+6)(N_{i^{mag}}+9)/162 $
$\alpha$-particles at the level $N_{i^{mag}}=3\nu$ (the $\nu$ is the principal
QN of the last occupied nucleon shell):

\begin{equation}
\Psi _A^{(SM)}=\hat A \prod_{i=1}^k \Psi_{\alpha_i}\cdot
\hat P \prod_{i=1}^k \varphi_{(3\nu0)}(R_i):\{(\lambda_i \mu_i) \}
[4^{A/4}] N \Omega L S T. \label{Mag}
\end{equation}
Naturally the values of QNs are $\{(\lambda \mu) \} =(00)$ $\Omega = L = S = T
= 0$ . Hence the WF of the ground state of the $^{40}$Ca nucleus can be
presented as the antisymmetrized product of ten $\alpha$-particles at the
$N_i=6$ level (the SU(3) $\alpha$-configuration $(60)^{10}$). Such a one-level
occupation picture is valid for a lot of MAPS of arbitrary N=Z-even nonmagic
nuclei, namely the ground state bands of the $^8$Be ($N_i=2$) and $^{20}$Ne
($N_i=4$) nuclei, the $N=9$ states of the $^{12}$C nucleus etc. So it is
possible for all $\alpha$-particles of a system to be concentrated at one and
the same level as it is inherent for the $\alpha$-condensate. However it cannot
be the lowest level and what is more the principal quantum number $N_i$ of this
level depends on the number of nucleons forming the system. The condition $N_i
\geq N_{i^{mag}}^>(4/3 - k_{mag}^>/3k)$ where $N_{i^{mag}}^>$ and $k_{mag}^>$
are the values of $N_i$ and $k$ for the nearest heavier magic nucleus, is
valid. Quite the reverse the allowed number of $\alpha$-particles $k$ at a
level depends on the quantum number $N_i$ of this level, in which case $k \leq
N_{i^{mag}}^> k_{mag}^> /(4N_{i^{mag}}^> - 3N_i)$. The value $k$ is by no means
the number of spinless one-particle orbital states at the oscillator level
$N_i$ which is equal to $(N_i+1)(N_i+2)/2$. Finally unlike the photon gas the
infinite number of $\alpha$-particles cannot concentrate at one level. The
statistics presented is neither the Bose-Einstein nor Fermi-Dirac one and even
not parastatistics.

Another principal peculiarity of the representations (\ref{C1}) and (\ref{Mag})
is that the occupation scheme is determined under the sign of the operators
$\hat P$ and $\hat A$. These operators preserve all global QNs ($N$ for
example) but the result of their operation (after the projecting with the use
of the operator $\hat O$) is a superposition of the functions (\ref{RSM}) with
various sets of the one-$\alpha$-particle QNs $N_i$. As a result the
occupations constructed in the schemes proposed are not selfreproducing. In
addition they are not unambiguous. Under the action of the operators $\hat P$
and $\hat A$ various occupation sets $\{N_i\}$ may result in one and the same
form of the WF $\Psi _A^{(SM)}$. Nevertheless the obtained from the given
procedures determine the main properties of MAPSs because the operator $\hat A$
preserves the global QNs. Moreover the statistics obtained is universal. This
means that irrespective of the Hamiltonian the MAP part of any solution (the
components which cannot be represented in MAP form appear in such a solution
for another nucleon Hamiltonian) can be expressed in terms of a superposition
of the components (\ref{Fun}) or, when multiplied by $\Phi_{000}(R_{cm})$, in
terms of a superposition of components (\ref{SM}) satisfying the selection
rules discussed.

It should be stressed that the statistics of arbitrary bosons composed by
fermions (mesons, atoms of isotopes with an even neutron number etc.) are
qualitatively the same as the $\alpha$-particle one. The number of bosons at a
certain level is limited due to antisymmetry of the WF of fermions comprising a
system. Thus only photons are the particles which rigorously obey the
Bose-Einstein statistics. So it seems to be reasonable to refer to this
property of the composite bosons as "quasi boson statistics" and to use for the
states discussed the term "quasi boson condensate" at least for systems
possessing an approximately equal size or a size which is few times greater
than that of comprising bosons.

The question arises concerning the behavior of the occupation numbers of
composite bosons with increasing of the system. The size of the $N_i$ orbital
$r$ has the form $r \simeq r_0 \sqrt{N_i+3/2}$ (the $r_0$ equals $\sqrt{\hbar
/m_{\alpha}\omega} $ for the $\alpha$-particle system) and the leading term
which determines the allowed number of the $\alpha$-particles $k$ at the level
$N_i$ is proportional to $N_i^3$. Thus the value $k$ increases as $k \sim
(r/r_{\alpha })^6$ and becomes in fact infinite for a system which is several
times larger than a nucleus. So the possibility to find a system which reveals
even a small deviation from the Bose-Einstein statistics other than the nucleus
is conjectural. Nevertheless a search for such deviation in small dense objects
consisting of large-size bosons seems to be very exciting.

It is also important to analyze the interrelation between the given model of
$\alpha$-condensation and the one proposed in the pioneer work \cite{TH}. The
definition of the latter is as follows:
\begin{equation}
\Psi_{cond} \equiv \hat A \prod_{i=1}^k \Psi_{\alpha_i} \hat P
\prod_{i=1}^k \exp[(-2/B^2)(R_i^2)]
, \label{Sch}
\end{equation}
where $B^2=b^2+2R_0^2$, the b, the oscillator parameter of its internal motion,
and $R_0$ is a measure of the size of the system as a whole.

The fuction $\Phi _{cond}(\{R_i\})= \prod_{i=1}^k \exp[(-2/B^2)(R_i^2)]$ in
(\ref{Sch}) can be written as a superposition of WFs of the basis (\ref{RSM})
(summation over QNs $N$, $(\lambda \mu)$ $\eta$). Consequently the state
discussed obeys the selection rules which were obtained above. The function
$\Phi _{cond}(\{R_i\})$ is selfreproducing only approximately because it
contains rather small (for large $R_0$) but nonzero forbidden components. The
operation of $\hat A$ on this function gives rise to some nodes in each
surviving component of the superposition.

Each of the exps. (\ref{Mag}) and (\ref{Sch}) describes the system of $k$
$\alpha$-particles which are located in the equivalent states of CM motion.
Thus the expressions are very similar even formally. At the same time the
present model has revealed a lot of condensed states different from the
low-density states described in \cite{TH}. In this sense the effect of
$\alpha$-particle condensation in some near-threshold $0^+$ states is a
particular case of a wide range of condensation effects in the various states
of different angular momentum. Obviously for some of the states (\ref{Mag}) an
energy level occupied by $\alpha$-particles turns out to be lower than for the
state (\ref{Sch}) in the same nucleus. And the lowest state corresponds to the
conventional condensate pattern to a greater extent.

Sometimes the $\alpha$-particle properties of such states are more pronounced
because the spectroscopic factors $W_\alpha$ (reduced widths $\gamma _\alpha $)
of the $\alpha$+target entrance channels for more dense systems are greater
than those for the systems of low density. The reason is that the overlap of
the WF of the $(k-1)\alpha$-particle subsystem of the low density
$k\alpha$-particle system with the WF of the ground state of the target nucleus
is small.

Consideration of $W_\alpha$ values sheds new light on the properties of low
density condensed states. Indeed, due to the smallness of the spectroscopic
factors the $\alpha$-decay widths of them should be rather small. From this
standpoint the width $\Gamma _\alpha$ = 4.8 MeV of the level of the $^{16}$O
14.0 MeV which is considered in \cite{TH} as $\alpha$-condensed state seems to
be too large because only $W_\alpha \sim 1$ may give such values of the widths.
Probably a more real candidate for the low-density condensed state in the
$^{16}$O nucleus is the 14.03 MeV level with $\Gamma _\alpha$ = 185 KeV.
Moreover the wide state 14.0 MeV has been rejected recently \cite{FS}.
According to this monograph the state 11.26 MeV with $\Gamma _\alpha$ = 2.6 MeV
which is also discussed in \cite{TH} is now open to question. It should be
noted that both the approach of \cite{TH} and the present one provide some
theoretical confirmation of this state. Indeed both the excitation energy
$E^*$=11.4 MeV and the value $W_\alpha=0.64$ of the $N=20, (\lambda \mu) =
(84), L=0$ state in the given model are in good agreement with experiment. And
what is more taking into account the fact that both approaches are valid for
the near-threshold states, one may conclude that in this state the overlap the
WF (\ref{Sch}) with the WF (\ref{Fun}) possessing the QNs just mentioned is
probably large. Indirect confirmation of this conclusion is a relatively small
root-mean-square radius of the WF (\ref{Sch}) $\sqrt{<r^2>} = 3.12$ fm (that is
why the value $W_{\alpha}$ is large). This radius is approximately the same as
that of WF (\ref{Fun}) with $N=20$. This fact is a further evidence of the
common features of the two approaches.

In conclusion we stress that the formalism developed based on SU(3) Hamiltonian
is in fact a precisely solvable model of the cluster stability of the certain
states. The qualitative picture of the processes in these states appears as
follows: $\alpha$-particles penetrate each other, and the interaction and
nucleon exchange result in disintegration (loss of individuality) of them and
yet the $\alpha$-particle properties of such a system are retained exactly.

The proposed multi-$\alpha$-particle model of a definite set of nuclear states
seems to be promising for investigating not only the problem of cluster
condensate in nuclei but also many other general problems concerning the
interrelation between the nucleon and cluster degrees of freedom.

Work supported by RFBR grant No.00-02-16683. The authors express their
gratitude to Profs. W.Scheid and I.Volobuev for fruitful discussions.

\end{document}